\documentclass[12pt,a4paper]{article}

\usepackage[utf8]{inputenc}
\usepackage[T1]{fontenc}
\usepackage{amsmath, amssymb}
\usepackage{graphicx}
\usepackage{booktabs}
\usepackage{longtable}
\usepackage{caption}
\usepackage{float}
\usepackage{natbib}
\usepackage{hyperref}

\usepackage[left=2.5cm,right=2.5cm,top=2.5cm,bottom=2.5cm]{geometry}
\usepackage{setspace}
\usepackage{authblk}

\title{Real-World fNIRS-Based Brain-Computer Interfaces: Benchmarking Deep Learning and Classical Models in Interactive Gaming}

\author[1,*]{Mohammad Ghalavand}
\author[1,+]{Javad Hatami}
\author[2,+]{Seyed Kamaledin Setarehdan}
\author[3,+]{Hananeh Ghalavand}
\affil[1]{Department of Psychology and Educational Sciences, University of Tehran, Tehran, Iran}
\affil[2]{School of Electrical and Computer Engineering, College of Engineering, University of Tehran, Tehran, Iran}
\affil[3]{School of Pharmacy, Shahid Beheshti University of Medical Sciences, Tehran, Iran}

\affil[*]{\texttt{mohammadghalavand@ut.ac.ir}}
\affil[+]{These authors contributed equally to this work}

\begin{document}

\maketitle

\begin{abstract}
Brain-Computer Interfaces (BCIs) enable direct communication between the brain and external systems, with functional Near-Infrared Spectroscopy (fNIRS) emerging as a portable and non-invasive method for capturing cerebral hemodynamics. This study investigates the classification of rest and task states during a realistic, interactive tennis simulation using fNIRS signals and a range of machine learning approaches. We benchmarked traditional classifiers based on engineered features, Long Short-Term Memory (LSTM) networks on raw time-series data, and Convolutional Neural Networks (CNNs) applied to Gramian Angular Field-transformed images. Ensemble models like Extra Trees and Gradient Boosting achieved accuracies above 97\%, while the ResNet-based CNN reached 95.0\% accuracy with a near-perfect AUC of 99.2\%, outperforming both LSTM and EfficientNet architectures. A novel data augmentation strategy was employed to equalize trial durations while preserving physiological integrity. Feature importance analyses revealed that both oxygenated and deoxygenated hemoglobin signals—particularly slope and RMS metrics—were key contributors to classification performance. These findings demonstrate the strong potential of fNIRS-based BCIs for deployment in dynamic, real-world environments and underscore the advantages of deep learning models in decoding complex neural signals.
\end{abstract}


\section{Introduction1}

The quest to establish direct communication pathways between the human brain and external devices has driven remarkable innovations in Brain-Computer Interfaces (BCIs) \citep{shenoy2006towards, naseer2015fnirs}. These systems represent not merely technological achievements, but transformative tools that fundamentally reshape our understanding of human-computer interaction. BCIs have demonstrated particular promise in enhancing independence for individuals with motor disabilities—enabling control of prosthetic limbs, wheelchairs, and communication systems \citep{bamdad2015application}—while simultaneously advancing our understanding of neural mechanisms underlying cognitive processes \citep{nijholt2008braingain, plass2010human}.

The effectiveness of any BCI system hinges upon its neuroimaging foundation. Contemporary systems leverage diverse modalities including electroencephalography (EEG) \citep{vaid2015eeg, jiang2024large, hameed2025enhancing}, magnetoencephalography (MEG) \citep{roy2019channel}, functional magnetic resonance imaging (fMRI) \citep{sorger2020real}, and functional Near-Infrared Spectroscopy (fNIRS) \citep{hong2015classification, zafar2023hybrid, ghaffar2021improving, pinti2020present, berivanlou2016quantifying}. Each technique presents distinct advantages and limitations in capturing neural signals that, once processed, can be translated into commands or used to classify specific mental states. The selection of an appropriate neuroimaging modality must consider factors including spatial and temporal resolution, portability, cost-effectiveness, and resilience to movement artifacts \citep{berivanlou2016quantifying, pinti2020present}.

Among these neuroimaging approaches, functional Near-Infrared Spectroscopy (fNIRS) has emerged as particularly promising for BCI applications. As a non-invasive optical technique, fNIRS measures changes in cerebral cortex oxygenated (HbO) and deoxygenated (HbR) hemoglobin concentrations \citep{pinti2020present, ferrari2012brief, hoshi2016hemodynamic, ghalavand2024comparison}. This measurement exploits neurovascular coupling—the physiological relationship where neural activity increases trigger corresponding rises in local blood flow to meet metabolic demands. This capability makes fNIRS particularly valuable for decoding sustained cognitive or motor states in BCI applications \citep{pinti2020present, hoshi2016hemodynamic}.

The growing adoption of fNIRS in BCI research stems from several practical advantages \citep{naseer2015fnirs}. Its portability enables brain activity monitoring in diverse environments beyond laboratory settings. The technology's relative ease of use and lower cost compared to alternatives like fMRI enhance accessibility for both research and potential clinical applications \citep{pinti2018review, pinti2015using, ferrari2012brief}. Furthermore, fNIRS demonstrates superior tolerance to motion artifacts compared to fMRI and EEG—a crucial advantage when investigating motor tasks or working with populations prone to movement. These benefits facilitate studies in more ecologically valid settings and with participant groups less suited to other neuroimaging methods. Nevertheless, fNIRS has important limitations, including restricted depth penetration (primarily measuring superficial cerebral cortex layers) and sensitivity to systemic physiological changes, particularly during physically demanding tasks \citep{pinti2018review, ferrari2012brief, naseer2015fnirs}.

A central challenge in cognitive neuroscience involves distinguishing neural signatures between different brain states—particularly between intrinsic resting activity and task-engaged states \citep{van2009tuning, broadbent2023cognitive}. Comparing these distinct activity patterns provides critical insights into the brain's functional organization and the neural mechanisms underlying various cognitive processes. This comparative approach forms the foundation for identifying brain regions and networks specifically involved in particular functions.

Simulation games, particularly tennis simulations, provide structured yet engaging paradigms for investigating brain activity related to control and cognitive engagement. Even in virtual environments, tennis play demands complex integration of visuomotor coordination, sustained attention, anticipation, and rapid strategic decision-making \citep{ghalavand2024comparison}. Using tennis simulation as an experimental task state offers the opportunity to examine brain activity under conditions approximating real-world cognitive and motor demands, potentially yielding more ecologically valid and generalizable findings.

In BCI development, machine learning algorithms serve as essential translators that convert complex fNIRS data streams into meaningful interpretations—classifying different brain states or predicting user intentions \citep{khan2021classification}. Researchers have applied diverse approaches to fNIRS data classification, from traditional methods like Support Vector Machines (SVM) \citep{jakkula2006tutorial} and k-nearest neighbors (KNN) \citep{guo2003knn}, to advanced deep learning \citep{lecun2015deep} architectures including Convolutional Neural Networks (CNN) \citep{kattenborn2021review} and Long Short-Term Memory networks (LSTM) \citep{yu2019review}. These algorithms identify characteristic patterns in fNIRS signals associated with different brain states, enabling BCIs to reliably differentiate between various cognitive and motor activities.

While existing literature establishes solid foundations for understanding fNIRS principles, BCI applications, and machine learning's role in brain state classification, a comprehensive comparison of classification algorithms specifically for distinguishing rest and task states during tennis simulation using fNIRS data remains underdeveloped. Understanding the comparative effectiveness of different classification approaches in this context is essential for optimizing fNIRS-based BCIs for interactive motor task applications and gaining deeper insights into the neural correlates of simulation engagement.

Accordingly, the present study aims to benchmark classical models, recurrent architectures, and CNNs applied to time-series and image-transformed fNIRS data. Our goal is to determine which computational strategy most effectively distinguishes between rest and cognitively demanding states during an interactive tennis simulation task—a scenario designed to mirror realistic cognitive-motor interplay.

\section{Materials and Methods}

\subsection{Participants}

In this study, 50 right-handed male participants with an average age of 24.94 years (SD = 3.77, ranging from 18 to 30) were included. Individuals with a history of psychiatric, neurological, or medical conditions, as well as those who had taken psycho-pharmacological medications within the last two months, were excluded. All participants had a basic knowledge of tennis, prior experience playing computer-based games, and no prior experience in professional tennis simulator gaming. The study received approval from the Research Ethics Committees of the Faculty of Psychology and Education (Approval ID: IR.UT.PSYEDU.REC.1401.074).

\begin{figure}[H]
\centering
\includegraphics[width=0.5\textwidth]{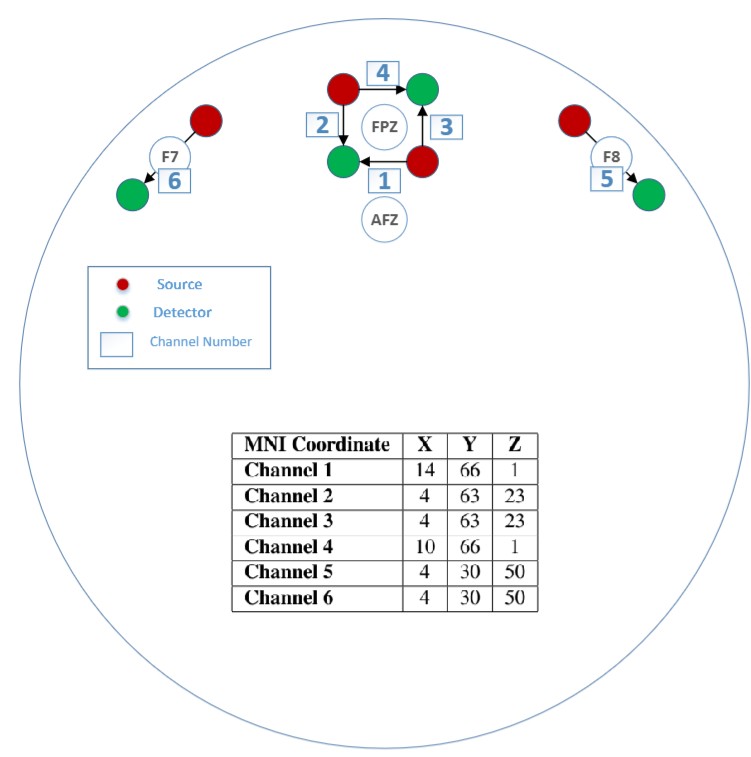}
\caption{Position of channels based on 10-20 system and MNI coordination}
\label{fig:fig1}
\end{figure}

\subsection{fNIRS Data Collection}

fNIRS data were collected using a 6-channel OxyMon system (Artinis Medical Systems). The system emitted light at wavelengths of 740 nm and 860 nm to measure the relative concentration of oxygenated and deoxygenated hemoglobin. Data were recorded at a sampling rate of 10 Hz. Optode placement followed the international 10-20 system, with a focus on the central medial prefrontal areas near FPz (Brodmann areas 9 and 10) and the lateral areas at F7 and F8 (Brodmann areas 47-45). The coordinates from the 10-20 system were converted to the MNI standard using fOLD software within the MATLAB environment as shown in figure.\ref{fig:fig1}

\subsection{Tennis Game as a Mind Reading Task}

\begin{figure}[H]
\centering
\includegraphics[width=0.8\textwidth]{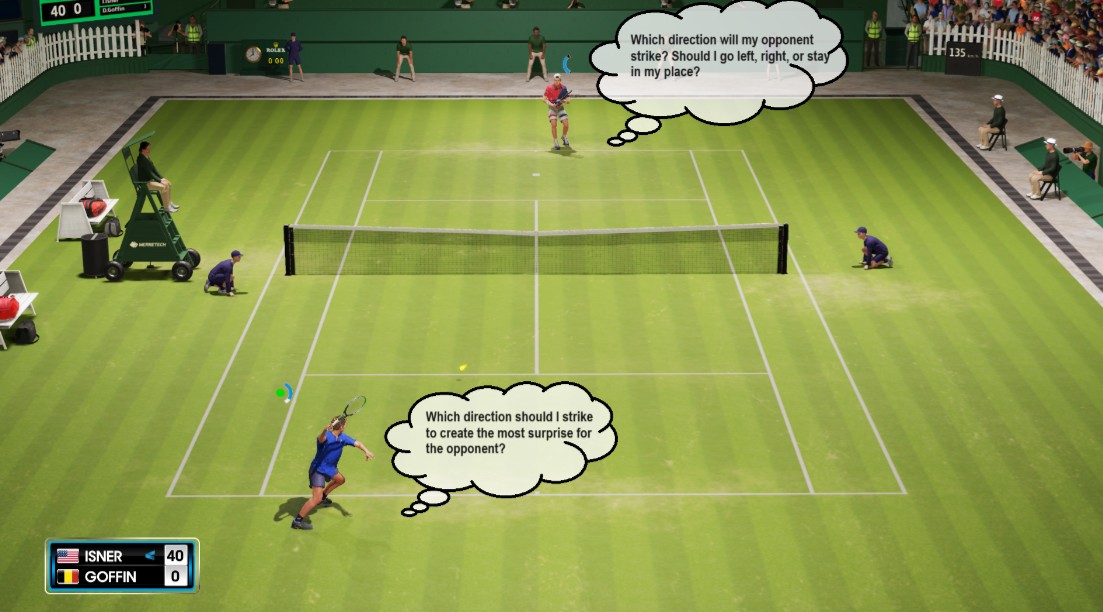}
\caption{The game environment. Participants must constantly anticipate the direction of their opponent's strike.}
\label{fig:fig2}
\end{figure}

A tennis simulation computer game (AO tennis 2) was used as an experimental task. The game allowed for single and two-player modes and adjustable difficulty levels. Participants were required to predict their opponent's actions and make rapid decisions. The game was designed with a block design and required continuous engagement of cognitive modules related to theory of mind.The game environment is shown in figure \ref{fig:fig2}.

\subsection{Experimental Paradigm}

During the experiment, participants took part in two 15-minute gaming sessions while undergoing fNIRS scanning. They played a simulated tennis video game under two conditions: (1) solo gameplay against an AI opponent and (2) two-player gameplay against a human opponent. Notably, although our previous paper \cite{ghalavand2024comparison} thoroughly examined the differences between artificial agent and human agent gameplay, the current study specifically investigates the contrast between activity and rest states under these continuous conditions.

To ensure proper timing, the experimenter gave verbal instructions. Participants were told to "stop" and close their eyes, then "continue" to resume gameplay. A timer was used to maintain accuracy, and the experimenter controlled the game remotely, pausing and restarting it at the beginning and end of each cycle.

In the solo condition, each participant (Participant 1 and Participant 2) played against a computer opponent. The session consisted of ten cycles: one minute of gameplay followed by 30 seconds of rest, totaling 15 minutes.

In the two-player condition, Participant 1 and Participant 2 competed against each other while being scanned simultaneously. The structure matched the solo condition, with alternating one-minute gameplay and 30-second rest periods, also lasting 15 minutes.

Participants were instructed to aim for victory in both conditions. In addition to their base compensation, they received performance-based rewards. In the human opponent condition, the final score determined the outcome. In the AI opponent condition, success depended on outperforming the computer’s score, creating an indirect competition.

\begin{figure}[H]
\centering
\includegraphics[width=0.8\textwidth]{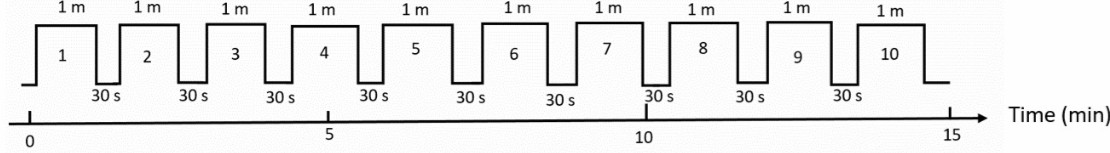}
\caption{Each game comprises ten cycles, with each cycle consisting of a 1-minute gameplay session followed by a 30-second rest period.}
\label{fig:fig3}
\end{figure}

To ensure consistent effort, participants were not informed of their final scores. Instead, the outcome was randomly determined by a coin flip after the sessions, encouraging full engagement in both human and AI gameplay.Figure \ref{fig:fig3} illustrates the experimental paradigm.

\subsection{Data Analysis}

The fNIRS data underwent preprocessing to enhance data quality and remove artifacts. The Beer-Lambert law was applied to obtain concentrations of oxygenated (oxyHb) and deoxygenated (deoxyHb) hemoglobin \citep{ferrari2012brief}. Datasets with noise, excessive movement artifacts, or insufficient signal quality were excluded, resulting in 45 samples for further analysis.

\subsection{Leveraging OxyHb and DeoxyHb Data to Improve fNIRS Interpretation}

In processing our fNIRS data, we rely on the principles of the Beer-Lambert law, which establishes a relationship between light absorption in a medium and the concentration of light-absorbing molecules like hemoglobin. This law serves as the theoretical foundation for measuring hemoglobin concentration changes, a key factor in assessing neural activity via fNIRS. Our research employed a system that automatically applies the Beer-Lambert law to raw optical signals, converting them into oxygenated (oxyHb) and deoxygenated (deoxyHb) hemoglobin concentration values.

Although prior studies have often restricted their analysis to oxyHb data, we have chosen to incorporate deoxyHb measurements as well. This approach follows the methodological recommendations outlined in Yücel et al.'s guidelines for fNIRS \citep{yucel2021best}. Including both oxyHb and deoxyHb is vital for a thorough assessment of neural dynamics, especially in cases where task-related physiological artifacts (e.g., altered breathing patterns during video gameplay could influence the results. By evaluating both hemoglobin species, we ensure a more complete representation of the hemodynamic response, allowing for deeper insights into neural activation under our experimental conditions \citep{tachtsidis2013investigation}.

\subsection{Preprocessing}

The preprocessing of functional near-infrared spectroscopy (fNIRS) data involves several critical steps to ensure optimal signal quality prior to machine learning or statistical analysis. The raw fNIRS data, initially stored in Excel files, undergoes a systematic transformation to enhance its suitability for downstream computational modeling.

First, the pipeline reads the raw optical signals from a specified input directory and applies a bandpass Butterworth filter (0.01–0.09 Hz) to isolate the hemodynamic frequency \citep{pinti2020present, tachtsidis2013investigation} range of interest while attenuating extraneous noise, including physiological artifacts such as cardiac pulsations and respiratory fluctuations. Following filtering, baseline correction is performed by subtracting the mean signal value, thereby normalizing the data and minimizing systemic drift.

Subsequently, the processed signals are segmented into distinct physiological states—active and rest conditions—based on predefined temporal intervals (e.g., category\_1\_ranges for task-active periods and category\_2\_ranges for rest periods). Each segmented time series is then exported as an individual CSV file, systematically organized into output directories with standardized filenames denoting their respective condition and source file.

This structured preprocessing pipeline converts raw fNIRS recordings into cleaned, labeled, and analysis-ready datasets, facilitating subsequent feature extraction or supervised model training.

Additionally In implementing a machine learning pipeline for classifying fNIRS signals, we encountered a fundamental issue: the mismatch in duration between rest and task recordings. Our rest data consisted of 30-second segments, while the task data spanned 60 seconds. For deep learning models uniform input lengths are essential, as they enable consistent batching and avoid complications associated with variable-length time series \citep{batista2004study, kaur2019systematic, bolboacua2023performance, wang2023influence}. Unequal input lengths not only complicate the architecture but also risk introducing padding-induced biases \citep{dwarampudi2019effects} or data truncation artifacts, which can degrade model performance. Therefore, ensuring that all inputs are of equal length was a necessary preprocessing step \citep{bolboacua2023performance, dwarampudi2019effects}.

One obvious approach to this problem is to truncate the longer (task) signals to match the shorter rest segments. While simple, this strategy sacrifices a significant portion of the task data, potentially discarding meaningful hemodynamic patterns related to cognitive activity. Especially in fNIRS research, where data collection is often limited due to participant availability and experimental constraints, reducing the dataset size further undermines the statistical power of the analysis and the generalizability of the trained models.

To address this issue without compromising data quantity, we developed a hybrid upsampling pipeline to extend the duration of the 30-second rest signals to 60 seconds. This approach combines two augmentation techniques: Fourier-based resampling \citep{rasche1999resampling} and nonlinear time-warping \citep{zhang2023warpformer}. Each rest trial is randomly assigned one of these methods per channel, allowing the transformed signals to retain realistic physiological variability. In addition, low-amplitude, band-limited noise is injected to simulate natural fluctuations in hemodynamic responses and prevent overfitting. Each resulting file is labeled with a tag ("\_F" for Fourier-based, "\_T" for time-warped) to indicate the applied transformation, enabling further analysis of augmentation-specific effects.

This strategy allowed us to preserve the entirety of the original dataset while standardizing input lengths across all trials. By opting for data augmentation rather than reduction, we enhanced model training with a more diverse and balanced dataset. Importantly, visual inspection and spectral analysis of the augmented signals confirmed that key fNIRS features—such as the canonical anti-correlation between oxyhemoglobin and deoxyhemoglobin—were preserved. This preprocessing step thus improved data compatibility with deep learning requirements without compromising physiological interpretability, ultimately supporting more robust and generalizable classification models.
\section{Benchmarking Classification Approaches for fNIRS Data: Feature-Based, LSTM, and CNN Performance}
Following the preprocessing stage, we employed a diverse set of machine learning models and approaches to comprehensively examine which methodology yields the highest accuracy in classifying cognitive states from our data. This comparative analysis was structured around three primary approaches:

\subsection{Feature Extraction and Classification with Classical and Neural Network}
Our classification methodology comprised two principal phases: feature engineering \cite{dong2018feature} and subsequent classification utilizing a diverse repertoire of machine learning algorithms, including both classical methods and a simple neural network. This approach was designed to systematically evaluate the efficacy of extracting salient characteristics from raw functional Near-Infrared Spectroscopy (fNIRS) time-series data and to comprehensively compare the performance of various modeling paradigms when operating on this derived feature set.

For each fNIRS recording, a comprehensive suite of time-domain and statistical features was meticulously extracted from the time series associated with each optode channel. This feature set encompassed fundamental descriptive statistics, including the mean, standard deviation, minimum, maximum, and median, which serve to characterize the central tendency, dispersion, and range of the physiological signal variations. To capture the temporal dynamics inherent in the fNIRS signals, we computed the slope of a linear regression fit to the time series, providing a measure of the overall trend, and the maximum absolute value of the first derivative, indicative of the peak rate of change. The Root Mean Square (RMS) was included as a robust measure of signal magnitude. Furthermore, higher-order statistical moments such as skewness and kurtosis were incorporated to describe the shape and symmetry of the signal's amplitude distribution. These features were concatenated across all channels to construct a unified, high-dimensional feature vector for each experimental recording.

The resulting feature vectors served as input to a diverse array of machine learning classifiers, selected to represent a broad spectrum of algorithmic paradigms commonly employed in biosignal processing. The selection included classical methods such as Support Vector Machines (SVMs), implemented with linear, Radial Basis Function (RBF), and polynomial (degree 3) kernels \cite{abdullah2021machine} to investigate the impact of different decision boundary complexities on classification performance. We also employed the instance-based K-Nearest Neighbors (k=5) algorithm \cite{zhang2017learning} and tree-based methods, specifically Decision Trees \cite{ying2015decision}, Random Forest \cite{salman2024random}, and Extra Trees \cite{geurts2006extremely}, which are known for their interpretability and ability to handle non-linear relationships. Ensemble learning techniques, such as Gradient Boosting \cite{natekin2013gradient} and AdaBoost \cite{schapire2013explaining}, were included to leverage the principle of combining multiple weak learners to achieve enhanced predictive accuracy and robustness. Linear models, including Logistic Regression \cite{feng2014robust}, configured with a maximum of 100 iterations to ensure convergence, and the Stochastic Gradient (SGDClassifier) \cite{jiang2021assessing}, were also utilized. Additionally, the probabilistic Gaussian Naive Bayes classifier \cite{yang2018implementation} was incorporated.

Recognizing the potential of neural networks in capturing complex patterns in high-dimensional data, we also included a simple feed-forward neural network model in our comparative analysis. This network consisted of an input layer receiving the feature vector, followed by two densely connected hidden layers with 64 and 32 nodes, respectively, both utilizing the Rectified Linear Unit (ReLU) activation function \cite{schmidt2020nonparametric,agarap2018deep}. The output layer comprised a single node with a sigmoid activation function for binary classification. The model was compiled using the Adam optimizer \cite{ahn2024understanding} and the binary cross-entropy loss function, with accuracy as the evaluation metric \cite{muller2016introduction}.

To obtain reliable and unbiased estimates of model performance and to mitigate the risk of overfitting, a 10-fold cross-validation strategy was systematically applied \cite{browne2000cross}. The dataset of feature vectors was partitioned into ten equally-sized subsets. In each iteration of the cross-validation procedure, one subset was reserved as the test set, while the remaining nine subsets were used for training the classifier. This process was iterated ten times, ensuring that each subset served as the test set exactly once. Model performance was quantitatively assessed using standard classification metrics: accuracy, precision, recall, and F1 score \cite{muller2016introduction}. These metrics provided a comprehensive evaluation of each model's discriminatory power on the feature-engineered fNIRS data, enabling a direct comparison of the different machine learning approaches based on their ability to classify fNIRS recordings into the predefined 'active' and 'rest' states using the extracted feature set.

\subsection{Evaluation of LSTM Performance on  Time-Series Data}
The proposed model is a deep learning architecture based on a Long Short-Term Memory (LSTM) network \cite{khan2024enhancing} enhanced with an attention mechanism \cite{bunterngchit2024temporal,liu2021research,liu2024news}, designed for multivariate time-series classification tasks. The input to the model is a 3D tensor with dimensions corresponding to time steps and the number of signal channels (features).

First, an LSTM layer with 64 units is employed to capture temporal dependencies in the input sequences. This layer is configured to return sequences, preserving the temporal structure for the subsequent attention mechanism. An attention block is then applied to emphasize the most informative time steps. The attention mechanism computes a set of attention weights across the temporal dimension, which are used to scale the LSTM outputs via element-wise multiplication. This process helps the model focus on critical moments in the input sequence that contribute more significantly to the classification task. .

Following the attention layer, the output is flattened and passed through a fully connected dense layer with 64 ReLU-activated units, followed by a dropout layer with a rate of 0.5 to reduce overfitting. The final output layer uses a softmax activation function \cite{pearce2021understanding}to perform multi-class classification.

The model is compiled with the Adam optimizer and categorical cross-entropy loss\cite{mao2023cross}. Performance metrics include accuracy, precision, recall, F1-score, and area under the ROC curve (AUC).

Data preprocessing includes normalization using the StandardScaler to ensure consistent scaling across features. For evaluation, stratified k-fold cross-validation is used to ensure balanced class representation in each fold. In each fold, early stopping is applied during training to prevent overfitting, with patience set to 10 epochs and the best weights restored based on validation loss. The results from all folds are aggregated and saved for subsequent analysis.

\subsection{Image Conversion and Classification using CNN Architectures}
In the third approach involves transforming the preprocessed raw fNIRS time-series data into a 2D image representation. This conversion enables the application of Convolutional Neural Network (CNN) architectures, specifically ResNet50 \cite{koonce2021resnet} and EfficientNet-B0 \cite{kansal2024resnet}, which are well-suited for exploiting spatial patterns in the data. This approach explores the potential of leveraging the spatial characteristics of fNIRS signals for enhanced classification accuracy. To overcome these limitations, a promising approach involves transforming fNIRS time-series data into two-dimensional (2D) image representations for subsequent analysis using Convolutional Neural Networks (CNNs). CNNs are particularly well-suited for image-based classification tasks due to their ability to automatically learn hierarchical features directly from the image data, thereby reducing the reliance on manual feature engineering. Representing fNIRS data as images can effectively capture both the temporal dynamics of the signals and the spatial relationships between different measurement channels \cite{park2023mental}. This transformation facilitates not only direct visual analysis but also enables the application of sophisticated image-driven machine learning techniques. Moreover, leveraging pre-trained CNN architectures, such as ResNet and EfficientNet, can lead to improved classification accuracy and a reduction in the time required for model training.

Transforming fNIRS time-series data from CSV files into visual representations using the Gramian Angular Field (GAF) technique\cite{wang2015imaging}. This process is applied to data stored in separate directories for 'active' and 'rest' conditions. For each CSV file representing fNIRS measurements, the data is first read and normalized to a range between -1 and 1 using a Min-Max Scaler. Following normalization, the Gramian Angular Field transformation is applied with a specified image size of 224 pixels, using the summation method to encode temporal correlations within the time series into a 2D image.

To leverage the multichannel nature of your fNIRS data (up to 12 channels), our method combines the individual GAF images generated for each channel into a single composite image. These individual channel GAF images are arranged in a 4x3 grid within the final RGB image. This approach allows for the visualization of the temporal dynamics of multiple fNIRS channels within a single spatial representation, potentially capturing both temporal and spatial characteristics of the brain activity related to the 'active' and 'rest' states. The resulting composite GAF images are then saved as PNG files in designated 'task' and 'rest' image directories, ready for subsequent analysis, such as with Convolutional Neural Networks designed to learn patterns from these image-based representations of fNIRS data.

Convolutional Neural Networks (CNNs) are highly effective for image classification, leveraging convolutional layers to extract spatial features, pooling layers to reduce dimensionality, and fully connected layers for final classification. Their ability to capture local patterns makes them well-suited for analyzing fNIRS-derived images, which encode brain activity data in 2D representations. To optimize performance, the CNN's input dimensions must match the fNIRS image size, and the number of input channels should align with the data structure (e.g., grayscale, HbO/HbR separation, or multi-sensor layouts). Transfer learning from models pre-trained on natural images (e.g., ImageNet) can enhance feature extraction, though careful fine-tuning is required to adapt to fNIRS-specific patterns.

Two prominent CNN architectures for fNIRS classification are ResNet and EfficientNet. ResNet's residual connections enable deep networks to detect subtle brain activity patterns while avoiding vanishing gradients. In contrast, EfficientNet prioritizes efficiency through depthwise separable convolutions, offering a lightweight alternative for resource-constrained environments, albeit with a potential trade-off in accuracy. Both architectures can be adapted to fNIRS data by adjusting input parameters and leveraging transfer learning.

The choice between ResNet and EfficientNet depends on the application's needs: ResNet excels in accuracy for complex fNIRS feature extraction, while EfficientNet is ideal for real-time or portable systems. Pre-processing steps—such as converting raw fNIRS time-series into 2D images (e.g., via Gramian Angular Fields or topography maps)—further influence model performance. By tailoring these components, CNNs can effectively classify fNIRS data for both research and clinical use.

\title{Results and Model Comparison}

\maketitle

\section{Results}

\subsection{Classical Models}

The performance of various machine learning models was evaluated using 10-fold cross-validation, with metrics including accuracy, precision, recall, F1 score, and ROC AUC. Among the tested classifiers, Gradient Boosting and Extra Trees achieved the highest mean accuracy (0.9725 ± 0.0132 and 0.9758 ± 0.0133, respectively), with Gradient Boosting also attaining the highest mean ROC AUC (0.9960 ± 0.0027). Linear SVM and Random Forest also demonstrated strong performance, with mean accuracies of 0.9599 ± 0.0128 and 0.9626 ± 0.0145, respectively.

Notably, Polynomial SVM, KNN, and Naive Bayes exhibited lower performance, with mean accuracies of 0.8440 ± 0.0367, 0.8313 ± 0.0347, and 0.8401 ± 0.0375, respectively. These models also showed higher variability across folds, particularly in precision and recall. Decision Tree performed moderately (mean accuracy = 0.8885 ± 0.0195), while Neural Networks and Logistic Regression achieved competitive results (mean accuracy = 0.9643 ± 0.0128 and 0.9648 ± 0.0108, respectively).

The RBF SVM classifier yielded a high mean ROC AUC (0.9926 ± 0.0053), though its accuracy (0.9544 ± 0.0224) was slightly lower than that of Linear SVM. AdaBoost and SGD classifiers performed comparably to ensemble methods, with AdaBoost achieving a mean accuracy of 0.9626 ± 0.0159.

Across all models, precision-recall trade-offs were observed: Polynomial SVM and KNN achieved near-perfect recall (0.9927 ± 0.0081 and 0.9945 ± 0.0055) but suffered in precision (0.7655 ± 0.0580 and 0.7497 ± 0.0514), suggesting a bias toward false positives. In contrast, Extra Trees and Gradient Boosting maintained balanced precision (0.9656 ± 0.0297 and 0.9592 ± 0.0287) and recall (0.9860 ± 0.0092 and 0.9857 ± 0.0079), indicating robust generalization.

\begin{figure}[H]
    \centering
    \includegraphics[width=0.8\textwidth]{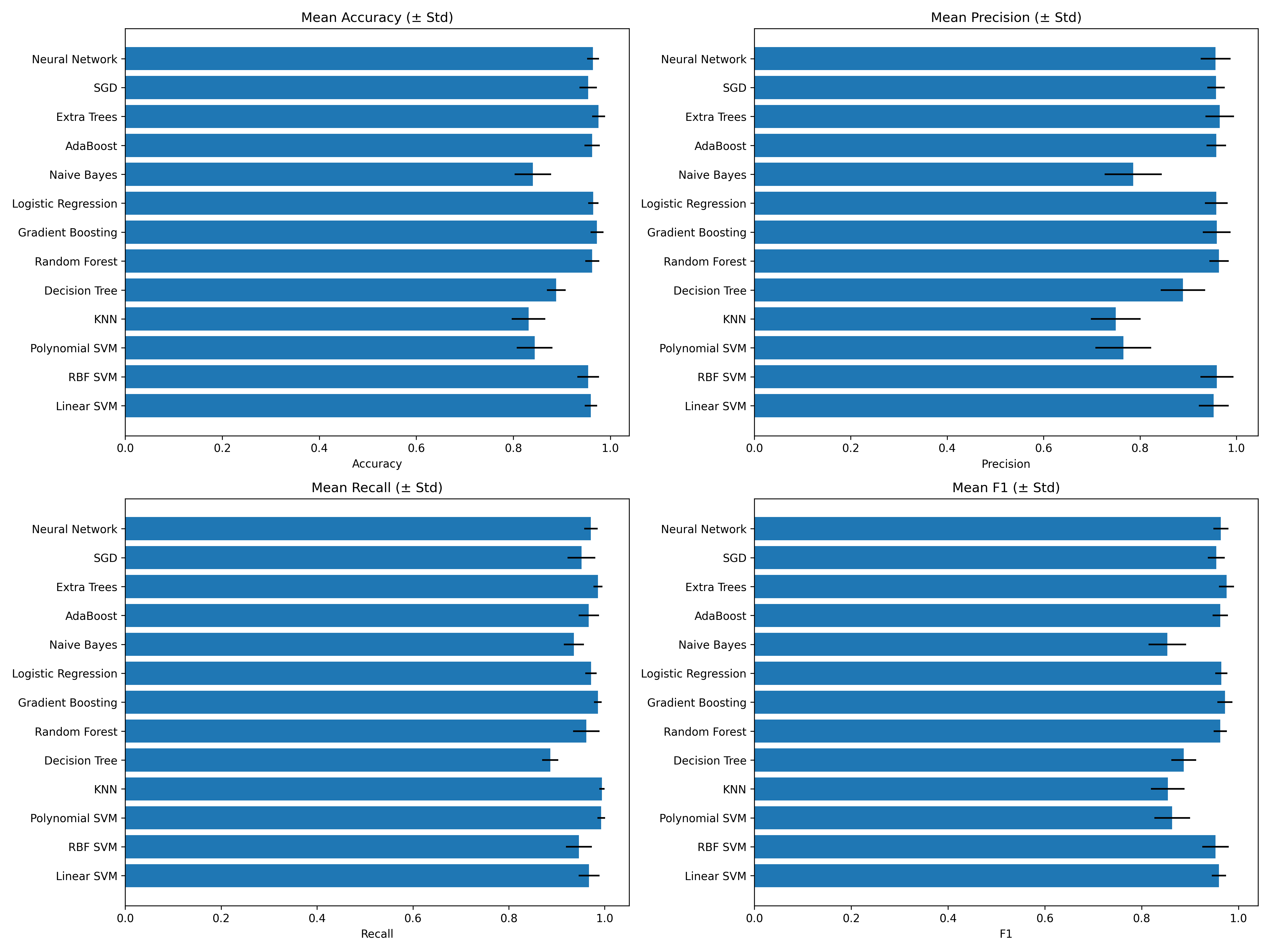}
    \caption{Performance of various machine learning models}
    \label{fig:myimage}
\end{figure}

The low standard deviations in ROC AUC for top-performing models (e.g., 0.0018 for Extra Trees) suggest stable discriminative ability, whereas higher variability in KNN and Naive Bayes highlights sensitivity to dataset partitioning.More details have shown in table1\ref{tab:scores} and figure\ref{fig:myimage}

\begin{table}[H]
\centering
\caption{Comparison of Traditional Machine Learning Models}
\label{tab:scores}

\resizebox{\textwidth}{!}{
\begin{tabular}{|l|c|c|c|c|c|c|c|c|c|}

\hline
\textbf{Model} & \textbf{Accuracy} & \textbf{Acc Std} & \textbf{Precision} & \textbf{Prec Std} & \textbf{Recall} & \textbf{Rec Std} & \textbf{F1} & \textbf{F1 Std} & \textbf{ROC AUC} \\
\hline
Linear SVM         & 0.95989  & 0.01278  & 0.953021  & 0.031034  & 0.967502  & 0.021718  & 0.959667  & 0.014714  & 0.991284 \\
RBF SVM            & 0.954396 & 0.022393 & 0.959604  & 0.034153  & 0.946309  & 0.026996  & 0.952703  & 0.027293  & 0.992593 \\
Polynomial SVM     & 0.843956 & 0.036711 & 0.765491  & 0.057976  & 0.992737  & 0.008073  & 0.863056  & 0.036854  & 0.950945 \\
KNN                & 0.831319 & 0.034668 & 0.749714  & 0.051410  & 0.994503  & 0.005512  & 0.853924  & 0.034928  & 0.942851 \\
Decision Tree      & 0.888462 & 0.019511 & 0.889017  & 0.046075  & 0.886582  & 0.016852  & 0.887118  & 0.025745  & 0.889051 \\
Random Forest      & 0.962637 & 0.014496 & 0.963921  & 0.020187  & 0.961883  & 0.027500  & 0.962477  & 0.013722  & 0.993850 \\
Gradient Boosting  & 0.972527 & 0.013233 & 0.959214  & 0.028731  & 0.985747  & 0.007919  & 0.972061  & 0.015511  & 0.996002 \\
Logistic Regression& 0.964835 & 0.010767 & 0.958212  & 0.023602  & 0.971391  & 0.011732  & 0.964559  & 0.012717  & 0.992318 \\
Naive Bayes        & 0.840110 & 0.037544 & 0.786005  & 0.059353  & 0.935919  & 0.021091  & 0.853180  & 0.038675  & 0.947096 \\
AdaBoost           & 0.962637 & 0.015887 & 0.958473  & 0.020354  & 0.966957  & 0.021430  & 0.962509  & 0.015945  & 0.992992 \\
Extra Trees        & 0.975824 & 0.013278 & 0.965555  & 0.029668  & 0.986015  & 0.009221  & 0.975400  & 0.015557  & 0.996876 \\
SGD                & 0.954396 & 0.018065 & 0.957593  & 0.018240  & 0.951634  & 0.028912  & 0.954302  & 0.017476  & 0.989985 \\
Neural Network     & 0.964286 & 0.012827 & 0.956636  & 0.027044  & 0.971093  & 0.013483  & 0.963578  & 0.015802  & 0.994946 \\
\hline
\end{tabular}
}
\end{table}

\subsection{LSTM Model}

The proposed LSTM-attention model demonstrated consistently strong performance across all 10 folds when applied to the fNIRS dataset. The mean classification accuracy reached approximately 93.1\%, with fold-specific accuracies ranging from 91.0\% to 97.4\%. Other key performance metrics—Precision, Recall, and F1-score—also indicated high reliability, with mean values of 94.2\%, 94.7\%, and 94.2\% respectively. Notably, several folds (e.g., folds 4, 5, and 6) exhibited near-ceiling performance across all metrics, suggesting the model’s ability to robustly capture underlying patterns associated with the cognitive or affective states in the fNIRS signals. Additionally, the Area Under the Curve (AUC) values remained consistently high across folds, all above 0.97, further emphasizing the model’s strong discriminative power.

Training convergence was also efficient, with most folds reaching optimal performance within 15 to 19 epochs, indicating stable and fast model optimization. The slight variability in metrics across folds may reflect individual or inter-session variability within the dataset, a common characteristic of neuroimaging data. Nonetheless, the consistently high performance across all folds supports the generalizability and robustness of the proposed architecture. Taken together, these results highlight the potential of attention-augmented LSTM networks for capturing the temporal dynamics of fNIRS data, thereby facilitating accurate classification of mental or behavioral states during dyadic interaction scenarios.

(For a detailed breakdown of results across folds, refer to Table~\ref{tab:scores2} )

\begin{table}[h]
\centering
\caption{Performance Metrics Across Folds}
\label{tab:scores2}
\begin{tabular}{cccccccc}
\toprule
\textbf{Fold} & \textbf{Accuracy} & \textbf{Precision} & \textbf{Recall} & \textbf{F1} & \textbf{AUC}  & \textbf{Best Epoch} \\
\midrule
1  & 0.910828025 & 0.896907216 & 0.956043956 & 0.925531915 & 0.97952048  & 7  \\
2  & 0.942675159 & 0.988095238 & 0.912087912 & 0.948571429 & 0.989177489 & 7  \\
3  & 0.923566879 & 0.954022989 & 0.912087912 & 0.93258427  & 0.985514486 & 6  \\
4  & 0.961783439 & 0.97752809  & 0.956043956 & 0.966666667 & 0.995504496 & 9  \\
5  & 0.955414013 & 0.977272727 & 0.945054945 & 0.960893855 & 0.99000999  & 9  \\
6  & 0.974358974 & 0.978021978 & 0.978021978 & 0.978021978 & 0.987489434 & 5  \\
7  & 0.955128205 & 0.9375      & 0.989010989 & 0.962566845 & 0.99509721  & 5  \\
8  & 0.929487179 & 0.944444444 & 0.934065934 & 0.939226519 & 0.984108199 & 8  \\
9  & 0.91025641  & 0.923076923 & 0.923076923 & 0.923076923 & 0.981403212 & 5  \\
10 & 0.942307692 & 0.945652174 & 0.956043956 & 0.950819672 & 0.988503804 & 8  \\
\bottomrule
\end{tabular}
\end{table}

\subsection{CNN Models}

\subsubsection{EfficientNet}
The proposed EfficientNet model demonstrated excellent performance across all evaluation metrics, achieving a mean accuracy of 92.3\% (range: 89.6--95.1\%) and a mean AUC of 97.8\% (range: 96.4--99.1\%). The model showed consistent results across all 10 folds, with precision and recall both averaging above 90\%, indicating a well-balanced classification performance. Notably, the best epoch typically occurred in the mid-to-late stages of training (epochs 14--41), suggesting stable learning without severe overfitting. These results highlight the model's robustness in handling fNIRS data's inherent variability and noise.

These results(see table \ref{tab:fficientNet}) position EfficientNet as a powerful deep learning solution for fNIRS signal classification tasks. The consistent high performance across metrics and folds demonstrates its potential for clinical and research applications where accurate classification is critical. Further optimization could enhance its suitability for real-time implementations while maintaining the demonstrated classification reliability.

\begin{table}[H]
\centering
\caption{EfficientNet model performance across 10 folds}
\label{tab:fficientNet}
\begin{tabular}{|c|c|c|c|c|c|c|c|}
\hline
\textbf{Fold} & \textbf{Accuracy} & \textbf{Precision} & \textbf{Recall} & \textbf{F1} & \textbf{AUC}  & \textbf{Best Epoch} \\
\hline
1 & 0.8956 & 0.8736 & 0.9048 & 0.8889 & 0.9642  & 19 \\
2 & 0.9231 & 0.9208 & 0.9394 & 0.9300 & 0.9743  & 14 \\
3 & 0.9011 & 0.8700 & 0.9457 & 0.9063 & 0.9760 &  16 \\
4 & 0.9505 & 0.9444 & 0.9551 & 0.9497 & 0.9848 &  41 \\
5 & 0.9176 & 0.8764 & 0.9512 & 0.9123 & 0.9826 & 26  \\
6 & 0.9066 & 0.8737 & 0.9432 & 0.9071 & 0.9735 & 27  \\
7 & 0.9396 & 0.9053 & 0.9773 & 0.9399 & 0.9853 & 29  \\
8 & 0.9066 & 0.9043 & 0.9140 & 0.9091 & 0.9669 & 24  \\
9 & 0.9451 & 0.9541 & 0.9541 & 0.9541 & 0.9910 & 21  \\
10 & 0.9451 & 0.9318 & 0.9535 & 0.9425 & 0.9832 & 44  \\
\hline
\end{tabular}
\end{table}

\subsubsection{ResNet-based fNIRS Classification Performance}
The ResNet architecture demonstrated exceptional classification performance across all evaluation metrics, achieving a mean accuracy of 95.0\% (range: 91.2--98.4\%) and a near-perfect mean AUC of 99.2\% (range: 98.1--99.9\%). The model maintained outstanding balance between precision (mean 94.5\%, range 89.6--98.8\%) and recall (mean 96.1\%, range 90.5--100\%), as reflected in the consistently high F1-scores (mean 95.5\%, range 90.5--98.2\%). Optimal performance was typically reached between epochs 16--21, indicating efficient convergence without overfitting despite training durations up to 28 epochs.

The ResNet model showed remarkable consistency across all 10 folds(see table\ref{tab:ResNet}, with particularly strong performance in later folds (Folds 9--10 achieving 96.7--98.4\% accuracy). The architecture's deep residual connections appear particularly well-suited for fNIRS data, handling both spatial patterns and potential signal variations effectively. Notably, the model achieved perfect recall (100\%) in Fold 9 while maintaining 94.8\% precision, demonstrating its ability to capture true positives without significant false positive trade-offs. The consistently high AUC scores (>98\% in all folds) confirm excellent discriminative capability.

\begin{table}[H]
\centering
\caption{ResNet50 model performance across 10 folds}
\label{tab:ResNet}
\begin{tabular}{|c|c|c|c|c|c|c|c|}
\hline
\textbf{Fold} & \textbf{Accuracy} & \textbf{Precision} & \textbf{Recall} & \textbf{F1} & \textbf{AUC}  & \textbf{Best Epoch} \\
\hline
1 & 0.9121 & 0.9048 & 0.9048 & 0.9048 & 0.9832 & 17 \\
2 & 0.9670 & 0.9697 & 0.9697 & 0.9697 & 0.9949 & 16 \\
3 & 0.9505 & 0.9368 & 0.9674 & 0.9519 & 0.9937 & 19 \\
4 & 0.9286 & 0.8958 & 0.9663 & 0.9297 & 0.9815 & 20 \\
5 & 0.9615 & 0.9747 & 0.9390 & 0.9565 & 0.9927 & 16 \\
6 & 0.9615 & 0.9551 & 0.9659 & 0.9605 & 0.9938 & 17 \\
7 & 0.9615 & 0.9355 & 0.9886 & 0.9613 & 0.9961 & 19 \\
8 & 0.9121 & 0.8969 & 0.9355 & 0.9158 & 0.9844 & 16 \\
9 & 0.9670 & 0.9478 & 1.0000 & 0.9732 & 0.9992 & 19 \\
10 & 0.9835 & 0.9882 & 0.9767 & 0.9825 & 0.9981 & 21 \\
\hline
\end{tabular}
\end{table}
\subsection{Interpretable Feature Selection: Leveraging SHAP, Model Coefficients, and Tree-Based Importance Measures }

In this study, various methods were employed to identify the most important features across different machine learning models, based on the internal structure and interpretability of each model. For more complex or non-linear models such as {Naive Bayes}, {Neural Network (Keras)}, {K-Nearest Neighbors (KNN)}, {RBF SVM}, and {Polynomial SVM}, we utilized the {SHAP (SHapley Additive exPlanations)} framework \cite{ponce2024practical,konig2021relative}. SHAP provides a unified measure of feature importance based on game theory, making it suitable for models that do not inherently expose feature weights or importances.

For {tree-based models} including {Decision Tree}, {Random Forest}, {Extra Trees}, {Gradient Boosting}, and {AdaBoost}, we extracted feature importances directly using the \texttt{feature\_importances\_} attribute \cite{rajbahadur2021impact,mishra2021survey} , which reflects the average contribution of each feature to the reduction in impurity across all trees. Meanwhile, for {linear models} such as {Linear SVM}, {Logistic Regression}, and {SGDClassifier}, we used the model coefficients (\texttt{coef\_}) as a measure of feature importance \cite{milner2015analysing}, since these weights indicate the strength and direction of the relationship between each feature and the predicted outcome.

The detailed results of these analyses are presented in the following sections.

\subsubsection{Coefficient-Based Feature Importance in fNIRS Classification }
This analysis utilized three distinct classification models—Stochastic Gradient Descent (SGD), Logistic Regression, and Linear Support Vector Classifier (LinearSVC)—to determine the importance of various features extracted from fNIRS signals. Across all models, feature importance was assessed using the coefficients assigned to each feature, where the magnitude and sign of these coefficients indicate the strength and direction of their contribution to the classification. A common thread in the findings was the significant role of features derived from both oxygenated hemoglobin (HbO) and deoxygenated hemoglobin (HbR) signals, particularly those related to root mean square (rms), maximum (max), and slope values. 

\begin{figure}[H]
\centering
\includegraphics[width=0.99\textwidth]{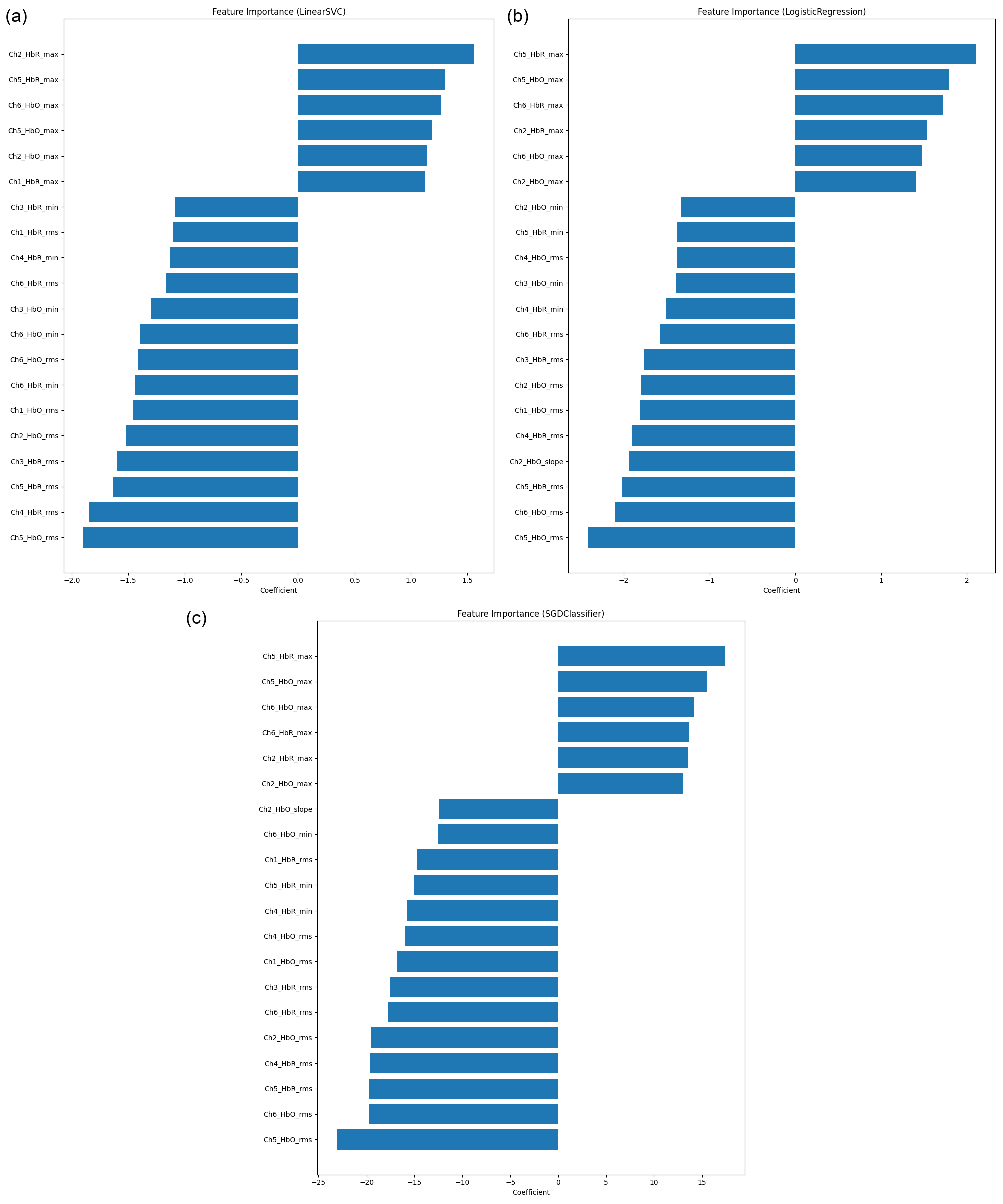}
\caption{a :LinearSVC , b :LogisticRegression c :SGDClassifier }
\label{fig:fig9}
\end{figure}

The results (see figure \ref{fig:fig9} ) from these analyses underscore the utility of coefficient-based feature importance for providing interpretable insights into which physiological signals are most predictive in fNIRS studies. The consistent influence of rms, max, and slope-based features suggests that both the amplitude and dynamic changes in hemodynamic responses are crucial for classification. Furthermore, the prominence of both HbO and HbR features among the top contributors indicates that a combination of both chromophores provides a more robust representation of brain activity. 

\subsubsection{SHAP Analysis for fNIRS Interpretability }
\begin{figure}[H]
\centering
\includegraphics[width=0.8\textwidth]{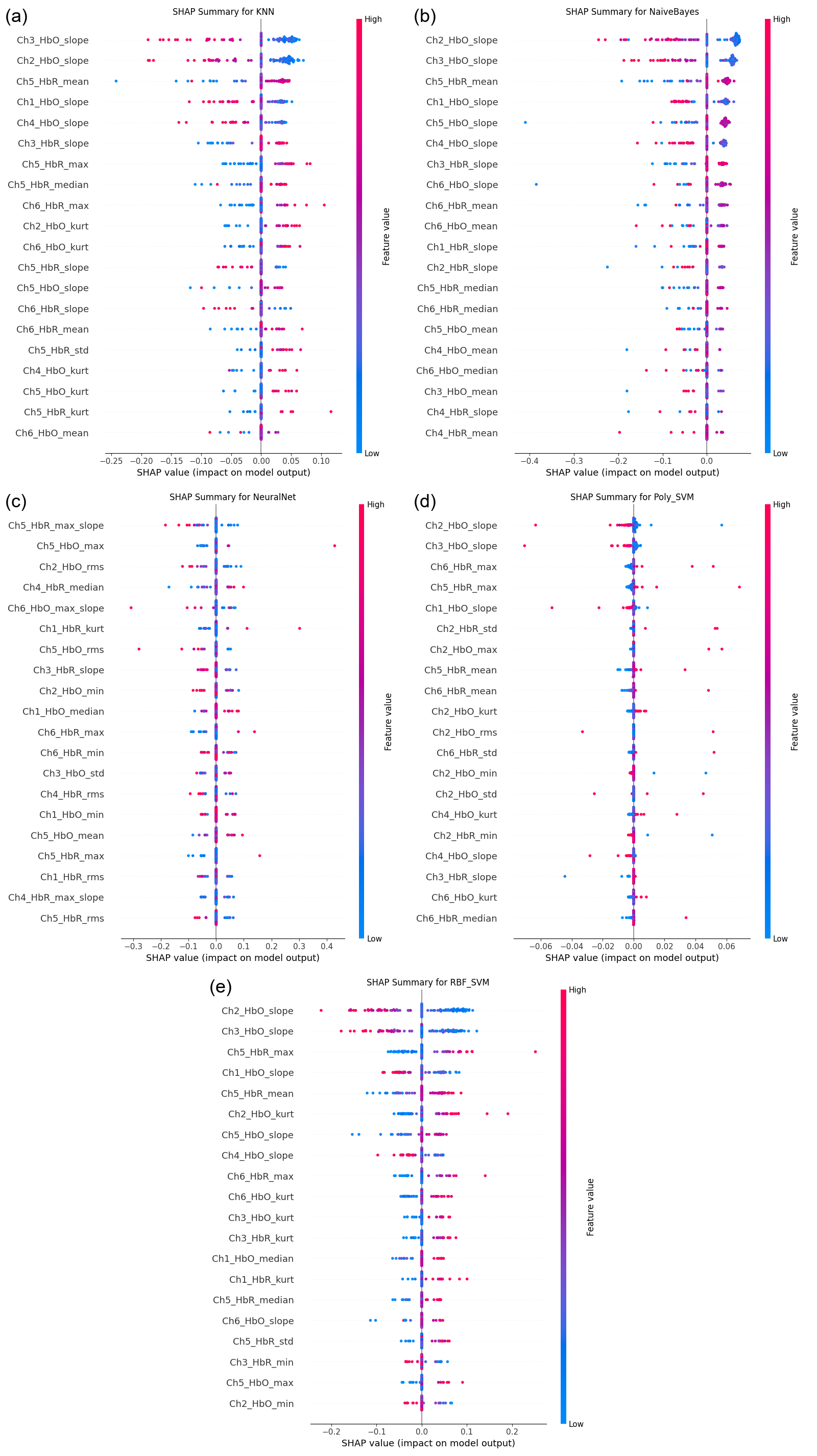}
\caption{a :KNN , b :NaiveBayes c :NeuralNetWORK d:PolySVM e :RbfSVM }
\label{fig:fig10}
\end{figure}

This analysis employed SHAP (SHapley Additive exPlanations) values to interpret the feature importance and contributions for five distinct fNIRS signal classification models: Polynomial SVM, RBF SVM, K-Nearest Neighbors (KNN), Naive Bayes, and a Neural Network. SHAP values provide a unified measure of feature importance, indicating how much each feature contributes to the model's prediction for individual instances, and collectively, their global importance. Across these diverse models, a consistent pattern emerged highlighting the significance of dynamic changes in hemoglobin concentrations, particularly features representing the slope of HbO signals from channels such as Ch1, Ch2, and Ch3. Additionally, statistical measures of HbR from channel 5, like its maximum or mean value, frequently appeared as influential. This suggests that both the rate of change in oxygenated hemoglobin and the magnitude of deoxygenated hemoglobin in specific frontal and parietal regions are crucial for distinguishing between experimental conditions.In figure \ref{fig:fig10}you can find more details.

The application of SHAP analysis provides robust and model-agnostic insights into the decision-making process of these fNIRS classifiers. Unlike linear coefficient methods, SHAP values can capture non-linear relationships and interaction effects, offering a more nuanced understanding of feature contributions. The general consensus across most models on the importance of specific HbO slope features and HbR magnitude features reinforces their physiological relevance in fNIRS-based classification tasks. The slight variations, particularly with the Neural Network, underscore how different model architectures might leverage feature information differently, yet still converge on broadly similar important physiological indicators. These detailed feature attributions are invaluable for validating model behavior and deepening the understanding of the neural correlates of the measured cognitive or behavioral states.
\subsubsection{Tree-Based Feature Importance }

This analysis evaluated feature importance for fNIRS signal classification using three tree-based ensemble models: Extra Trees, Gradient Boosting, and Random Forest. These methods assess feature importance by quantifying how much each feature contributes to reducing impurity (e.g., Gini impurity or entropy) across the decision trees in the ensemble, or by measuring the permutation importance. A striking consistency was observed across all three models, with features representing the slope of oxygenated hemoglobin (HbO) from channels 2 and 3  consistently ranking as the most influential. This indicates that the rate of change in HbO in these specific frontal channels is a primary driver for classification by these models.

\begin{figure}[H]
\centering
\includegraphics[width=0.99\textwidth]{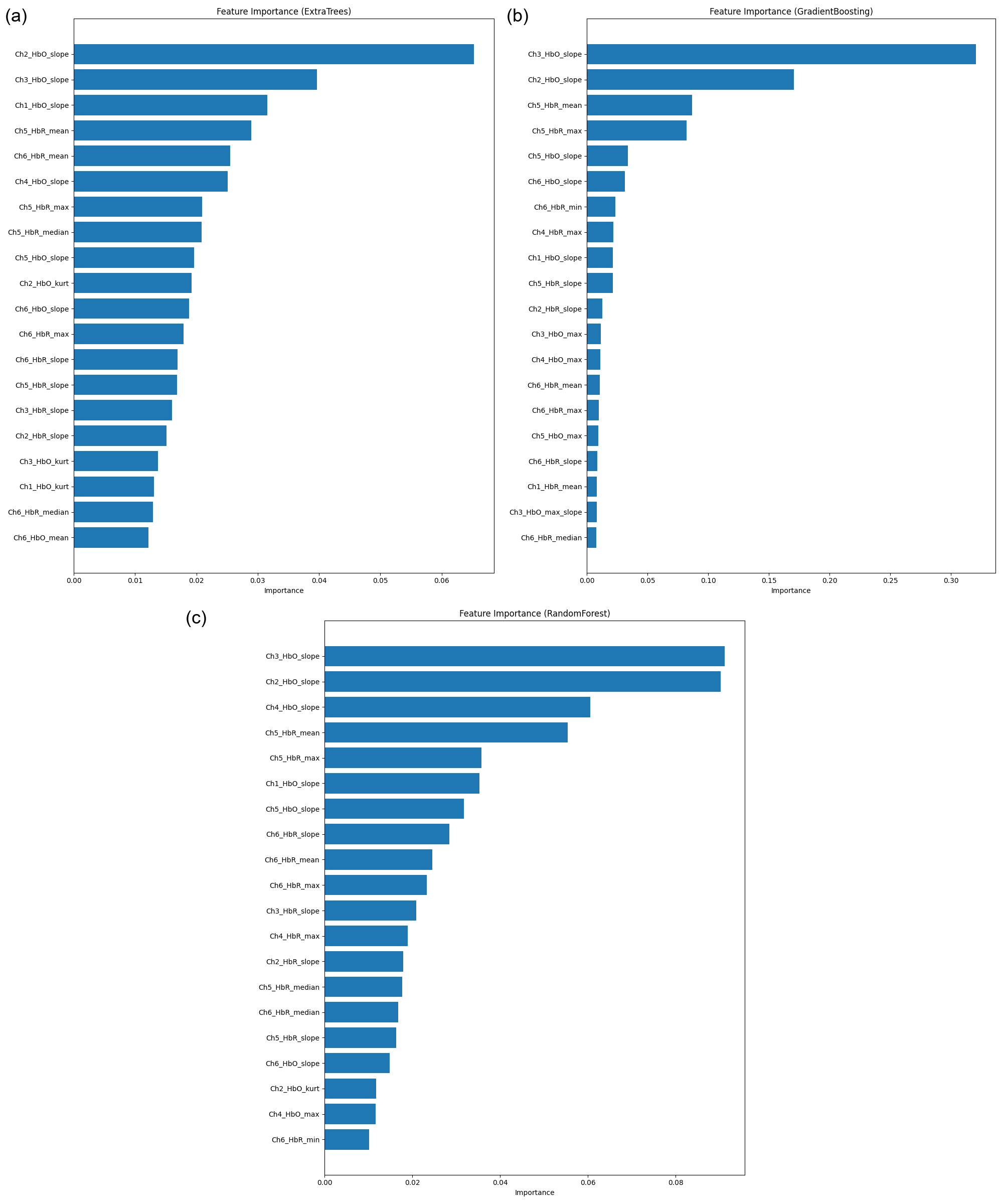}
\caption{a :ExtraTrees , b :GradientBoosting c :RandomForest }
\label{fig:fig11}
\end{figure}

The collective results from these tree-based ensemble methods strongly emphasize the predictive power of dynamic changes in HbO, particularly its slope, within frontal brain regions (channels 1, 2, 3, and 4). Additionally, measures of HbR magnitude, such as mean and maximum values, from channels 5 and 6, also play a significant, albeit generally secondary(see figure \ref{fig:fig11} ), role. The high degree of agreement between Extra Trees, Gradient Boosting, and Random Forest regarding the top-ranked features instills confidence in the robustness of these findings. This suggests that these specific fNIRS-derived features are reliable indicators for the classification task at hand, providing valuable insights into the underlying neurophysiological changes being captured.

\section*{Discussion}

This study provides a systematic comparison of machine learning methodologies for classifying rest and task states from fNIRS signals in a realistic, interactive gaming environment. The findings underscore the high discriminative power of fNIRS signals when processed with appropriate computational strategies and demonstrate that deep learning approaches—particularly Convolutional Neural Networks (CNNs)—can effectively harness the complex temporal and spatial dynamics embedded in these data.

\subsection*{Comparison of Model Performance}

Among the various machine learning paradigms evaluated, \textbf{ensemble-based classical models} such as Extra Trees and Gradient Boosting emerged as strong performers, achieving accuracies exceeding 97\% and AUC scores approaching 0.997. These results highlight the effectiveness of feature engineering in conjunction with robust tree-based algorithms for fNIRS classification, confirming the utility of statistical and temporal descriptors in capturing the underlying hemodynamic responses.

However, \textbf{deep learning models}—especially CNNs trained on Gramian Angular Field (GAF)-transformed images—demonstrated superior performance in several key areas. The ResNet architecture achieved a mean accuracy of 95\% and an AUC of 99.2\%, outperforming both EfficientNet and the LSTM-based recurrent network. ResNet’s deep residual connections likely contributed to its ability to generalize well despite the inherent noise and variability in fNIRS data. By contrast, EfficientNet achieved slightly lower accuracy (\textasciitilde92.3\%) but maintained high AUC and strong precision-recall balance, making it a compelling choice for resource-constrained applications.

The \textbf{LSTM-attention model} achieved a commendable average accuracy of 93.1\%, underscoring the value of temporal modeling in capturing the sequential dependencies of hemodynamic fluctuations. However, its performance lagged slightly behind CNNs, possibly due to challenges in modeling complex spatial interactions across multiple channels, which are more naturally captured in 2D representations.

\subsection*{Physiological Insights and Feature Interpretability}

Feature importance analyses across models consistently emphasized the relevance of both oxygenated (HbO) and deoxygenated (HbR) hemoglobin measures. Features such as slope, root mean square (RMS), and maximum values emerged as key contributors across coefficient-based, tree-based, and SHAP interpretability frameworks. These results affirm that not only the magnitude but also the \textbf{dynamics of hemodynamic responses}—particularly in frontal areas (channels 1–4)—are critical indicators of cognitive engagement.

Moreover, the consistent prominence of both HbO and HbR features across interpretable models supports recent methodological recommendations to consider both chromophores in fNIRS analyses. This dual-channel approach likely mitigated artifacts stemming from systemic physiological confounds and provided a more comprehensive depiction of cortical activation.

\subsection*{Methodological Innovations and Contributions}

A noteworthy contribution of this study lies in the \textbf{data augmentation strategy} employed to balance rest and task segment durations. Rather than truncating valuable task data, a hybrid augmentation pipeline incorporating Fourier resampling and nonlinear time-warping was implemented to upsample rest segments. This innovation preserved temporal characteristics and physiological interpretability, enhancing model performance without sacrificing data integrity.

Furthermore, the \textbf{use of GAF image transformation} represents an important methodological advance in fNIRS-BCI research. By converting time series into structured image representations, this approach enabled the application of pre-trained CNNs and capitalized on spatial hierarchies in the data, thereby improving classification accuracy and robustness.

\subsection*{Ecological Validity and Real-World Implications}

This study’s design emphasizes \textbf{ecological validity} by employing a dynamic, interactive gaming scenario—a tennis simulation—that demands visuomotor coordination, anticipation, and decision-making. The use of such a task bridges the gap between traditional laboratory paradigms and real-world cognitive-motor activities, enhancing the relevance of the findings for practical BCI applications. The high classification accuracy achieved across models confirms the feasibility of using fNIRS-based BCIs in dynamic environments, including neuroadaptive gaming, rehabilitation, and real-time cognitive monitoring.

\subsection*{Robustness Check: Evaluating Model Performance on Strictly Unseen Participants}
It may also be argued that since the data of each individual in the test set is likely also present during training, the results may not replicate when tested on new individuals, and the high accuracy in some models may not be due to learning the actual patterns related to rest and task. For this reason, we completely separated the data of 10 individuals in each fold and then trained the model on the remaining 35 individuals. In this case, the results remained nearly the same as before, indicating that the model indeed learns rest- and task-related patterns independently of the individuals. The results have been shown in the table \ref{tab:model_performance}  for this different approach:

\begin{table}[h]
\centering
\caption{Average of Models Performance Metrics for Completely Separated Situations}
\label{tab:model_performance}
\begin{tabular}{lcccc}
\toprule
\textbf{Model} & \textbf{Accuracy} & \textbf{Precision} & \textbf{Recall} & \textbf{F1 Score} \\
\midrule
Linear SVM & 0.925 & 0.920792079 & 0.93 & 0.925373134 \\
RBF SVM & 0.905 & 0.897058824 & 0.915 & 0.905940594 \\
Polynomial SVM & 0.78 & 0.697183099 & 0.99 & 0.818181818 \\
KNN & 0.79 & 0.707142857 & 0.99 & 0.825 \\
Decision Tree & 0.8775 & 0.891191710 & 0.86 & 0.875318066 \\
Random Forest & 0.9375 & 0.935323383 & 0.94 & 0.93765586 \\
Gradient Boosting & 0.9525 & 0.941463415 & 0.965 & 0.95308642 \\
Logistic Regression & 0.925 & 0.920792079 & 0.93 & 0.925373134 \\
Naive Bayes & 0.81 & 0.756198347 & 0.915 & 0.828054299 \\
AdaBoost & 0.9625 & 0.951219512 & 0.975 & 0.962962963 \\
Extra Trees & 0.9625 & 0.955665025 & 0.97 & 0.962779156 \\
SGD & 0.9375 & 0.958115183 & 0.915 & 0.936061381 \\
Neural Network & 0.945 & 0.927884615 & 0.965 & 0.946078431 \\
\bottomrule
\end{tabular}
\end{table}
\subsection{Methodological Reflections on Signal Duration and Data Augmentation}

A key methodological consideration in this study was managing unequal durations between rest and task states in the fNIRS recordings. Since deep learning models require uniform input lengths, we extended the 30-second rest-state segments to 60 seconds using a hybrid upsampling approach, combining Fourier-based resampling and nonlinear time warping.

While this allowed for consistency in training, it raised an important question: is the model learning from real physiological patterns, or could synthetic augmentation be contributing to its performance?

To address this, we tested two hypotheses:

\begin{enumerate}
    \item \textbf{Shorter task windows may not fully capture the Hemodynamic Response Function (HRF)}, leading to poorer classification.
    \item \textbf{Synthetic augmentation alone does not add discriminative value} unless anchored in real signal features.
\end{enumerate}

\begin{figure}[h]
\centering
\includegraphics[width=0.8\textwidth]{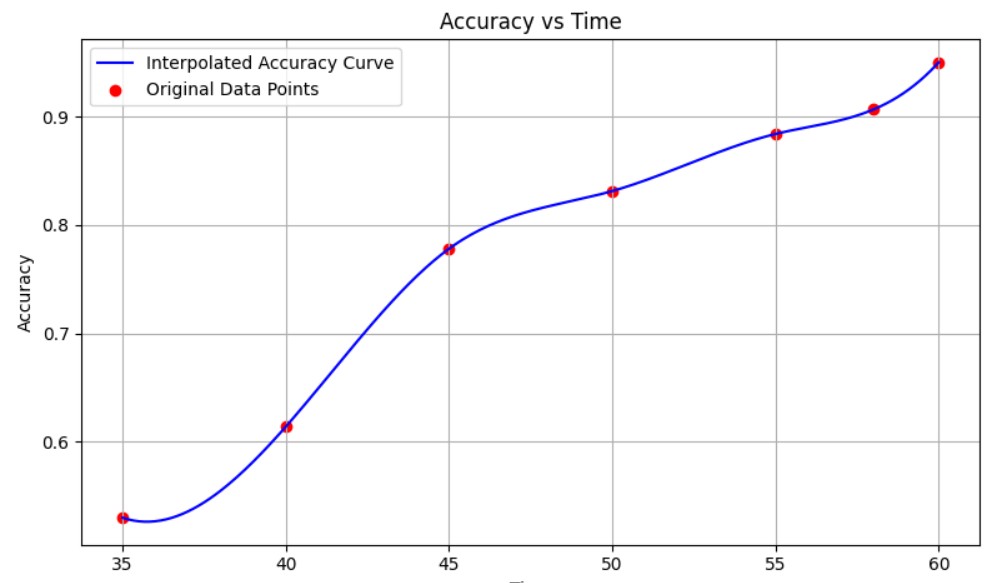}
\caption{Effect of gameplay window duration on model accuracy. As the task duration decreases, accuracy drops, suggesting the importance of full HRF dynamics.}
\label{fig:duration_accuracy}
\end{figure}

To evaluate these, we ran two experiments:

\begin{itemize}
    \item \textbf{Progressive truncation of task segments} from 60 seconds to 35 seconds(for Resnet50). As shown in Figure~\ref{fig:duration_accuracy}, accuracy dropped steadily as duration decreased, supporting the importance of full HRF capture.
    
    \item \textbf{Synthetic extension of short (30s) task segments}. Performance on these was substantially lower than on full-length originals, showing that augmentation alone cannot recover lost information.
\end{itemize}

\subsubsection*{Additional Validation with Mixed Real+Synthetic Input}

To test the limits of synthetic data's usefulness, we created a new dataset where each input segment (rest and task) consisted of 30 seconds of real data and 30 seconds of synthetically upsampled data. As shown in Table~\ref{tab:hybrid_augmentation_results}, this hybrid setup resulted in dramatic drops in model performance.

\begin{table}[h]
\centering
\caption{Performance when using 30s real + 30s synthetic input for both rest and task classes.}
\label{tab:hybrid_augmentation_results}
\begin{tabular}{|c|c|c|c|c|c|}
\hline
\textbf{Fold} & \textbf{Accuracy} & \textbf{Precision} & \textbf{Recall} & \textbf{F1 Score} & \textbf{AUC} \\
\hline
1 & 0.467 & 0.453 & 0.750 & 0.565 & 0.477 \\
2 & 0.531 & 0.543 & 0.889 & 0.674 & 0.447 \\
3 & 0.495 & 0.500 & 0.174 & 0.258 & 0.473 \\
4 & 0.489 & 0.489 & 1.000 & 0.657 & 0.469 \\
5 & 0.522 & 0.477 & 0.634 & 0.545 & 0.532 \\
6 & 0.478 & 0.431 & 0.250 & 0.317 & 0.498 \\
7 & 0.473 & 0.477 & 0.943 & 0.634 & 0.355 \\
8 & 0.511 & 0.514 & 0.763 & 0.615 & 0.503 \\
9 & 0.581 & 0.599 & 1.000 & 0.749 & 0.542 \\
10 & 0.527 & 0.500 & 0.093 & 0.157 & 0.490 \\
\hline
\end{tabular}
\end{table}

These results confirm that synthetic data—while useful for structural balancing—cannot replace real, temporally rich neural signals. When real signal content is diluted or replaced, the model’s learning ability diminishes significantly, underscoring the importance of physiological validity in BCI datasets.

\subsection*{Challenges and Future Perspectives}
Despite the promising results demonstrated in this study, several challenges must be addressed to further the practical deployment of fNIRS-based BCIs in real-world applications. One primary limitation lies in the intrinsic properties of fNIRS technology itself. While fNIRS is portable and resistant to movement artifacts, it remains limited in spatial resolution and depth penetration, only capturing activity from superficial cortical layers. This constraint may exclude critical brain regions involved in deeper cognitive processes, limiting the system's versatility for broader applications.

Another challenge pertains to inter-subject variability in hemodynamic responses. Even with sophisticated preprocessing and normalization techniques, individual differences in cortical anatomy, scalp thickness, and physiological baseline can introduce noise and reduce generalizability. Although our leave-subject-out validation demonstrated robustness, further studies with larger and more diverse participant cohorts are needed to confirm cross-population applicability.

Data augmentation strategies—especially those used to match segment lengths—while effective in preserving dataset size, also raise concerns regarding the authenticity of synthesized signals. Although spectral analyses confirmed physiological plausibility, synthetic signals may still lack the nuanced temporal dynamics present in genuine neural activity. Careful validation is necessary to ensure that augmentation does not inadvertently introduce biases or overfitting.

From a computational perspective, while deep learning models like CNNs (particularly ResNet) offered high performance, they demand significant computational resources, potentially limiting real-time deployment on portable or wearable systems. Additionally, model interpretability remains a key hurdle, especially for clinical or neurorehabilitative use, where transparent decision-making is essential.

Future research should focus on several fronts. First, integrating multimodal neuroimaging (e.g., combining fNIRS with EEG) could overcome modality-specific weaknesses and provide a richer picture of brain activity. Second, developing adaptive algorithms that personalize classification models based on user-specific neurophysiological profiles could enhance both accuracy and usability. Finally, improving the interpretability of deep learning models through explainable AI (XAI) frameworks will be critical to build trust and ensure safe, accountable deployment in sensitive settings.

\section*{Conclusion}
This study systematically benchmarked classical and deep learning models for classifying cognitive states from fNIRS signals during an ecologically valid tennis simulation game. Our findings reveal that both traditional machine learning approaches and modern deep neural architectures can effectively differentiate between rest and task-engaged brain states when properly preprocessed and structured.

Among the tested methods, deep CNNs—particularly ResNet trained on Gramian Angular Field-transformed data—achieved the highest classification performance, underscoring the power of spatially-aware representations in decoding neural signals. Classical ensemble models like Extra Trees and Gradient Boosting also delivered competitive accuracy, highlighting the continued value of feature engineering when data is limited.

Key contributions of this work include the development of a physiologically grounded data augmentation pipeline, the demonstration of robust model performance even on unseen participants, and the use of advanced interpretability tools to elucidate the neural features driving classification decisions. Importantly, this study advances the ecological relevance of fNIRS-based BCI research by applying machine learning in a realistic gaming context that closely mimics naturalistic cognitive-motor demands.

Overall, the results support the viability of fNIRS as a practical and scalable neuroimaging modality for brain-computer interfaces. With continued innovation in signal processing, model development, and deployment optimization, fNIRS-BCIs have strong potential for future use in adaptive gaming, cognitive monitoring, neurofeedback, and rehabilitation technologies.

\bibliographystyle{plainnat}
\bibliography{references}
\end{document}